\documentclass[12pt,preprint,a4paper]{aastex}

\usepackage{graphics,epsf}
\usepackage{amsmath}                
\usepackage{amsfonts}               
\usepackage{amssymb}                
\usepackage{epsfig}                 

\title{Planets around extreme horizontal branch stars}

\author{Ealeal Bear\altaffilmark{1} and Noam Soker\altaffilmark{1}}

\altaffiltext{1}{Department of Physics, Technion $-$ Israel
Institute of Technology, Haifa 32000 Israel; ealealbh@gmail.com;
soker@physics.technion.ac.il.}

\begin{document}

\begin{abstract}
We review three main results of our recent study:
\begin{itemize}
\item We show that a proper treatment of the tidal interaction
prior to the onset of the common envelope (CE) leads to an enhance
mass loss. This might increase the survivability of planets and
brown dwarfs that enter a CE phase. \item From the distribution of
planets around main sequence stars, we conclude that around many
sdB/sdO stars more than one planet might be present. One of these
might have a close orbit and the others at about orbital periods
of years or more. \item We show that the intense ionizing flux of
the extreme horizontal branch star might evaporate large
quantities of a very close surviving substellar object. Balmer
emission lines from the evaporated gas can be detected via their
Doppler shifts.
\end{itemize}
\end{abstract}


\section{Introduction}
\label{sec:Sec0}

EHB (Extreme Horizontal Branch) stars are hot, small, helium
burning stars. In order to become EHB the RGB progenitor must lose
most of its envelope. For the purpose of these proceedings there
will be no differentiation between sdB, sdO (which are the
spectroscopical classes) and EHB stars (the photometric
definition). It is known that planets can exist around sdB stars.
In many cases it is possible that planets are responsible for the
formation of the EHB star, e.g. HD 149382 b (Geier et al. 2009).
In these proceeding we will review three main topics: In section
1, we will discuss the interaction of brown dwarfs or low main
sequence stars and the RGB progenitor. Our goal is to study in
more detail the evolution of binary systems in a stage prior to
the onset of the CE phase, and in particular systems that have
reached synchronization; the synchronization is between the
orbital period and the primary rotation period. In section 2 we
will discuss the bimodal distribution of planets, and the
implications to EHB stars. In section 3 we will discuss planet
evaporation, and detection of H$\alpha$, and H$\beta$ emission. In
section 4 we will summarize our concussions.

\section{Interaction between brown dwarfs or low main sequence stars and the RGB progenitor}
\label{sec:Sec1}
We start our calculation when the primary stellar radius has
increased enough for tidal interaction to become significant. For
the binary systems we study, where the primary is an RGB star and
the secondary is a low-mass main sequence (MS) star or a brown
dwarf, tidal interaction becomes important when the giant swells
to a radius of $R_g \sim 0.2a$, where a is the orbital separation
and $R_g$ is the giant radius (Soker 1998). The primary radius
increases along the RGB as the core mass increases. The primary
and the secondary initial masses range in the binary systems we
study are $0.8M_\odot \leq M_1\leq 2.2M_\odot$, and
$0.015M_\odot\leq M_2\leq 0.2M_\odot$, respectively.
\begin{figure}
  \includegraphics[height=.5\textheight]{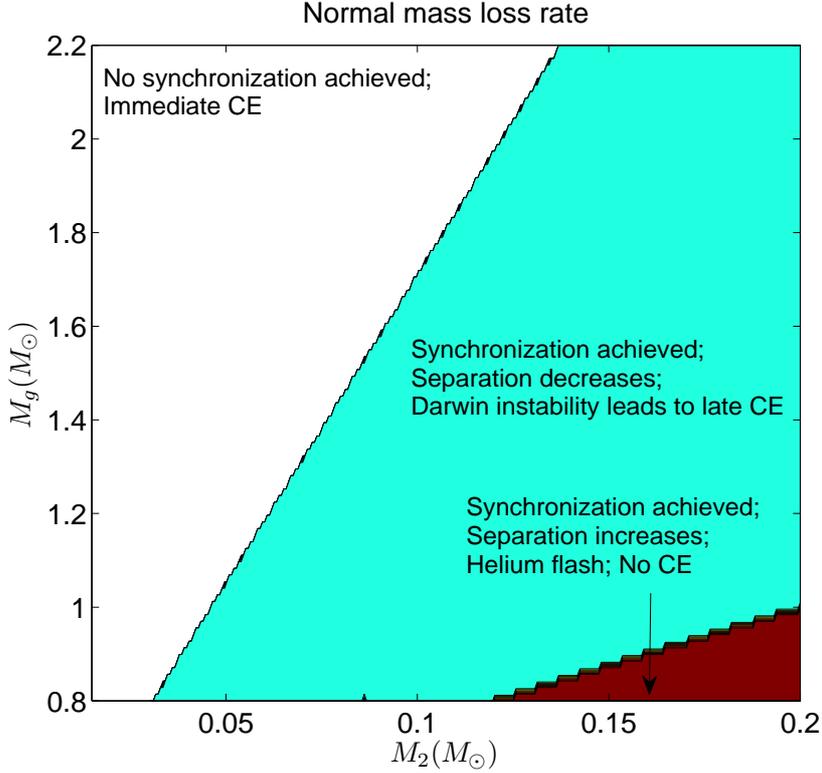}
  \caption{$M_g$ vs. $M_2$ representing binary systems that reached or did not reach synchronization.
Our normal mass-loss rate of $\eta_R = 3 \times 10^{14}$ was used
(for details see Reimers 1975, $\&$ Bear \& Soker 2010a. The
initial core mass is $Mc(0) = 0.4M_\odot$, and the initial
(primordial) orbital separation prior to tidal interaction is $a_0
= 5R_g(0) = 405R_\odot$. Calculation is terminated as indicated,
when one of the following occurs: The Darwin instability brings
the system to a CE phase (marked Darwin Instability); The core
mass reaches $0.48M_\odot$ and assumed to go through a core Helium
flash (Helium flash). The two other channels:
   (1) A total depletion of the RGB envelope.
    (2) The RGB stellar radius exceeds the orbital
separation ($R_g = a$), do not occur for the case presented
here.}\label{sbdf3}
\end{figure}

When a binary system starts its evolution it is not synchronized,
and therefore tidal interaction will lead to a fast spiraling-in
process. The binary system can then either reach a synchronization
or stays asynchronous. In systems that maintain synchronization
two opposing forces act: On one hand in order to maintain
synchronization the secondary transfers, via tidal forces, orbital
angular momentum to the envelope, and the orbit shrinks. On the
other hand mass loss acts to increase orbital separation. If
orbital separation increases faster than the RGB stellar radius,
no Common Envelope (CE) will occur either due to total envelope
loss or to a Helium flash. However, if orbital separation
decreases relative to the RGB stellar radius, and He flash or
total envelope loss do not occur too early the secondary enters
the envelope either due to Darwin instability, or by the swelling
RGB envelope. By that time the envelope mass is lower the the
initial envelope mass. These processes are represented in figure
\ref{sbdf3}.

As can be seen from the figure tidal interaction before the
formation of the CE increases the likelihood of low mass
companions to survive the CE phase. Furthermore, higher mass loss
rate decreases the chance of a late CE formation, but if late CE
occurs, it does so with lower envelope mass (for details see Bear
\& Soker 2010a).

\section{Bimodality of planets around MS stars}
\label{sec:Sec2}

\begin{figure}
  \includegraphics[width=0.9\textwidth]{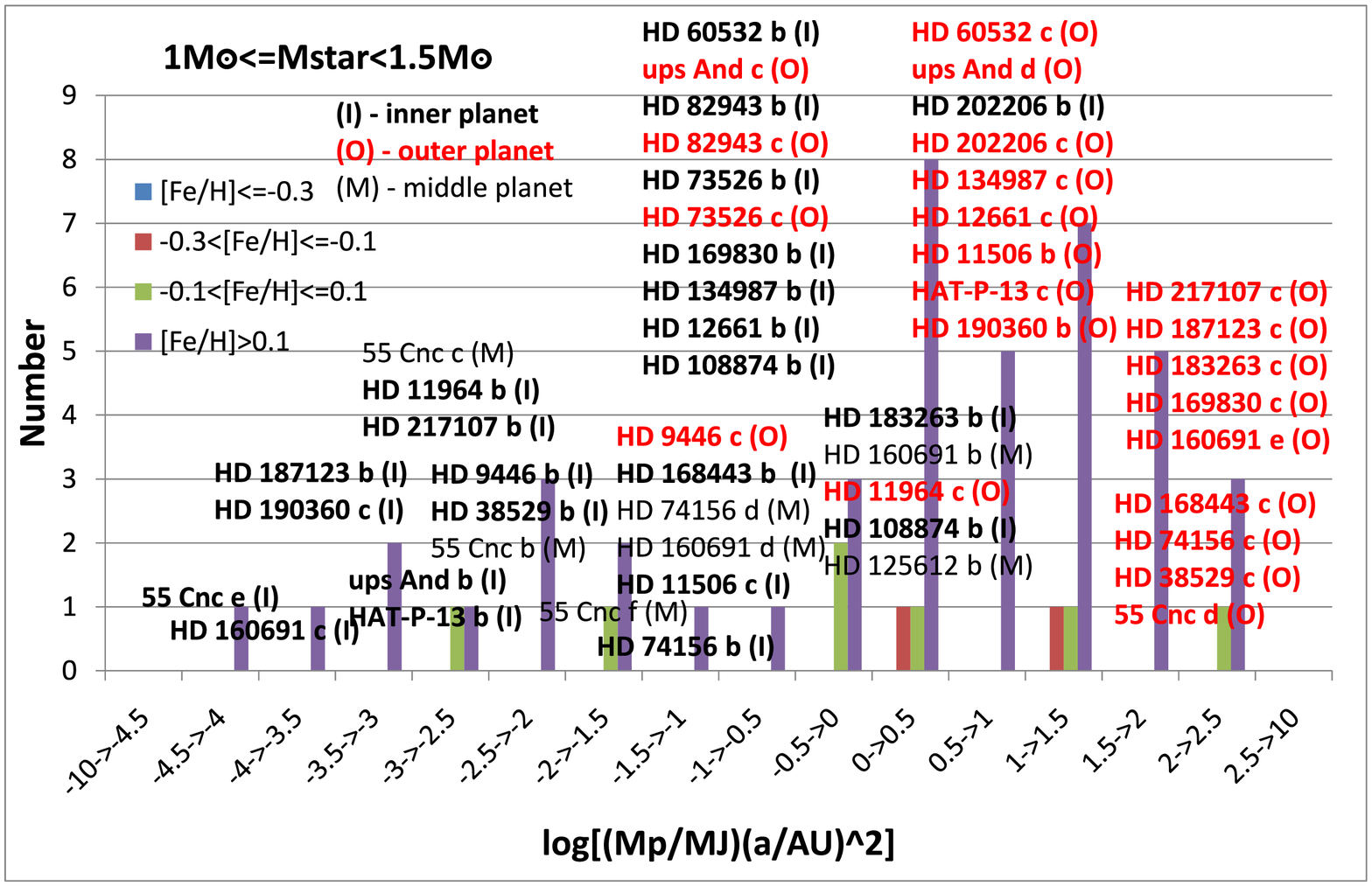}
  \caption{Number of planets vs. log $\left[\frac{M_p }{M_J}{\left(\frac{a}{AU}\right)}^2\right] $ . The planets
shown here are multi planet systems for stars in the mass range of
$1M_\odot \leq M_{\rm star} \leq 1.5M_\odot$.} \label{Multi_2}
\end{figure}
\begin{figure}
  \includegraphics[width=0.9\textwidth]{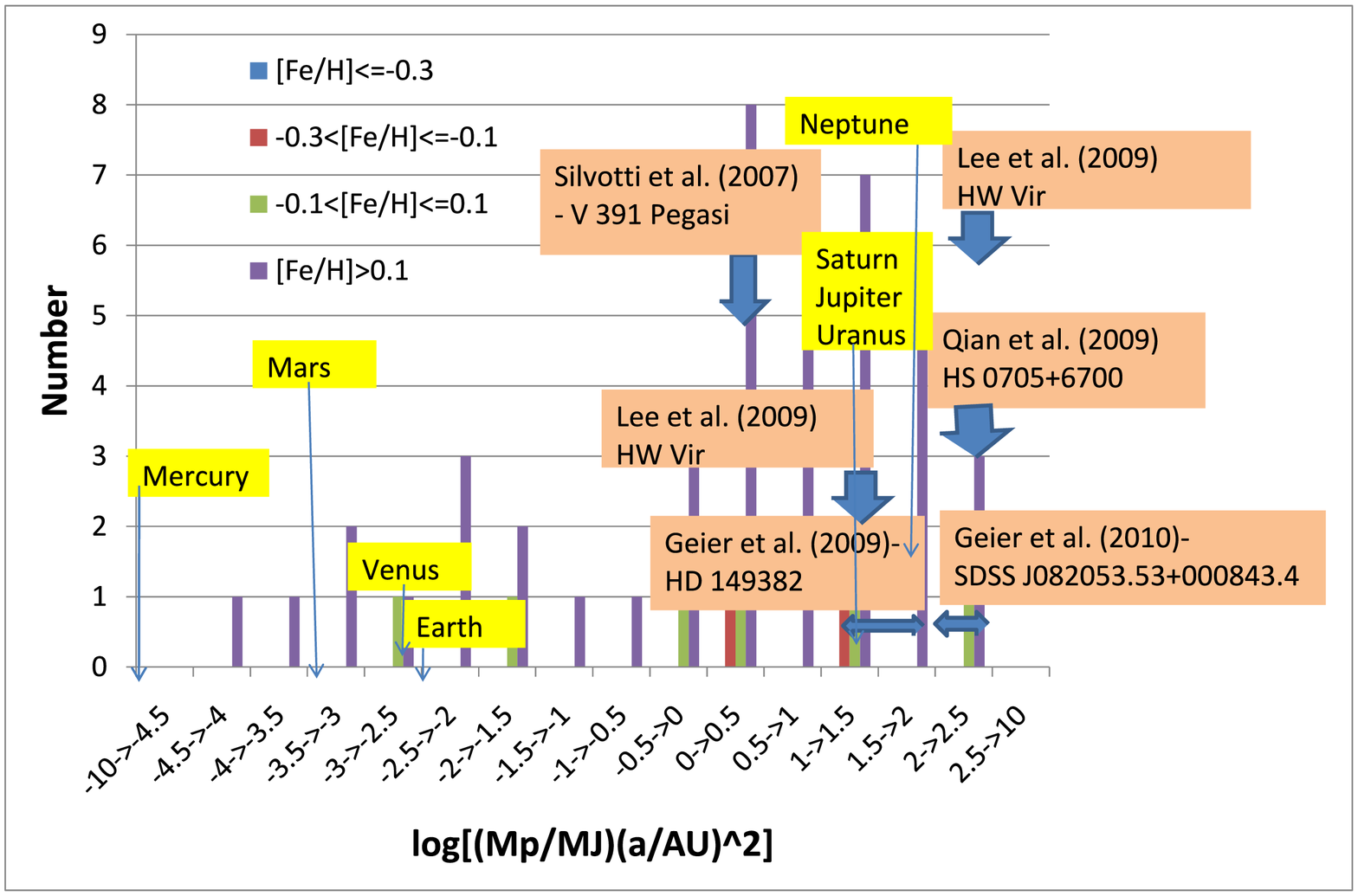}
  \caption{Number of planets vs. log $\left[\frac{M_p }{M_J}{\left(\frac{a}{AU}\right)}^2\right] $ . The planets
shown here are multi planet systems for stars in the mass range of
$1M_\odot \leq M_{\rm star} \leq 1.5M_\odot$, plus the planets of
our solar system.} \label{Multi_2_var1}
\end{figure}

We study the distribution of exoplanets around main sequence (MS)
stars and apply our results to the binary model for the formation
of extreme horizontal branch (EHB; sdO; sdB; hot subdwarfs) stars.
By the binary model we refer not only to stellar companions to RGB
stars (Han et al. 2002; Han et al. 2003), but to substellar
objects as well (Soker 1998). The bimodal distribution presented
in figure \ref{Multi_2} is taken from the Planets Encyclopedia
edited by Jean Schneider; http://exoplanet.eu/).

In this work we follow Soker \& Hershenhorn (2007). Soker \&
Hershenhorn (2007) examined the number of planets as a function of
metallicity bins and the planet mass $M_p$, orbital separation
$a$, and orbital eccentricity $e$, in several combinations. They
found that planets orbiting high metallicity stars tend to part
into two groups in a more distinct way than planets orbiting low
metallicity stars. They also found that high metallicity stars
tend on average to harbor closer planets. Soker \& Hershenhorn
(2007) had 207 planets in their analysis. We repeated their
analysis using 331 planets (out of more than 400 that were
discovered so far) and got similar results. In figure
\ref{Multi_2} only the multi-planet systems from the research are
represented (for more details see Bear \& Soker 2010b).

The planets shown in this figure are all part of a multi-planet
systems (two planets or more that orbit the same star). All
central stars are in the mass range of $1 - 1.5 M_\odot$, this
mass is the typical mass of the progenitors of EHB stars (for more
details see Bear \& Soker 2010b). In general, there are two groups
of planets, the inner marked by (I), and the outer planets that
are marked by (O). In order to implement this distribution to EHB
stars we will review the observations of planets around EHB stars:

Geier et al. (2009) announced recently the discovery of a close
substellar companion to the hot subdwarf (EHB) star HD 149382. The
orbital period is very short, 2.391 days, implying that the
substellar companion had evolved inside the bloated envelope of
the progenitor RGB star (a CE phase). The mass of the companion is
$8 - 23M_J$ , so either it is a planet or a low mass brown dwarf.
This discovery supports the prediction of Soker (1998) that such
planets can survive the common envelope (CE) phase, and more
relevant to us, that planets can enhance the mass loss rate on the
RGB and lead to the formation of EHB.

However, Jacobs et al. (2010) analyzed He lines and found no
evidence for the presence of this claimed planets. This debate
will soon be resolved by further observations. Other planets that
orbit EHB at larger separations have been detected (Silvotti et
al. 2007; Lee et al. 2009 \& Qian et al. 2009). Silvotti et al.
(2007) announced the detection of a planet with a mass of $3.2M_J$
and an orbital separation of $1.7 AU$, around the hot subdwarf
V391 Pegasi. Serendipitous discoveries of two substellar
companions around the eclipsing sdB binary HW Vir at distances of
$3.6AU$ and $5.3AU$ (Lee et al. 2009) and one brown dwarf around
the similar system HS 0705+6700 with a period of $2610 d $ and a
separation of $< 3.6AU$ (Qian et al. 2009). Recently Geier et al.
(2010) discovered  a brown dwarf companion to the hot subdwarf
SDSS J083053.53+0000843.4. This system contains an sdB star with
an approximated mass of $0.25-0.47M_\odot$. A brown dwarf of
$0.045-0.067M_\odot$ orbits this sdB with an orbital period of
$0.096{\rm d}$. Due to its close orbit it is very likely that this
system went through a Common Envelope phase.

All of these five systems are present in the graph, with their
evaluated orbital separation around the progenitor of the EHB star
(Bear \& Soker 2010b). It is quite plausible that closer planets
did interact with the RGB progenitor of the sdB star; they are not
observed in these systems. We end by noting that all these
substellar companions have been detected in the field. The main
conclusions we can draw from figure \ref{Multi_2_var1} are as
follows: Planets are expected in a double peak distribution. Outer
planets can survive the evolution. In particular, inner planets in
the same system that were engulfed, saved the outer planets by
enhancing mass loss rate early on the RGB. The inner planet might
survive the CE phase and be found around the EHB star, but only if
massive enough $M \geq 10 M_J$.

\section{Planet evaporation and detection}\label{sec:Sec3}


We study the evaporation of planets orbiting EHB stars. We adopt
the simple model presented by Lecavelier des Etangs (2007) which
represents the blow-off mechanism (Erkaev et al. 2007) and
investigate the implications for a planet orbiting an HB star
(this model is similar to the energy limited model purposed by
Murray-Clay et al. (2009). We refer to the ionization model as
well as a lower limit for the mass loss from the planet (for
details see McCray \& Lin 1994).

When the central source is hot a large fraction of the radiation
is energetic enough to ionize the evaporated gas. The evaporated
gas recombines and emits at longer wavelength, a radiation that
escapes from the planet's vicinity. Although recombination is not
relevant to planet around solar-like stars, its role becomes more
important for hot HB stars and central stars of planetary nebulae.
We assume that:
\begin{itemize}
\item Most of the evaporated gas flows toward the radiation
source. \item The central star keeps the gas almost fully ionized.
\item The ionizing photons of the parent star that are absorbed by
the evaporated gas are removed from the radiation that heat the
star. \item All the radiation emitted by the recombining gas
escape. \item We assume that the gas flows with the escape
velocity from the planet. \end{itemize}

\begin{figure}
  \includegraphics[height=.5\textheight]{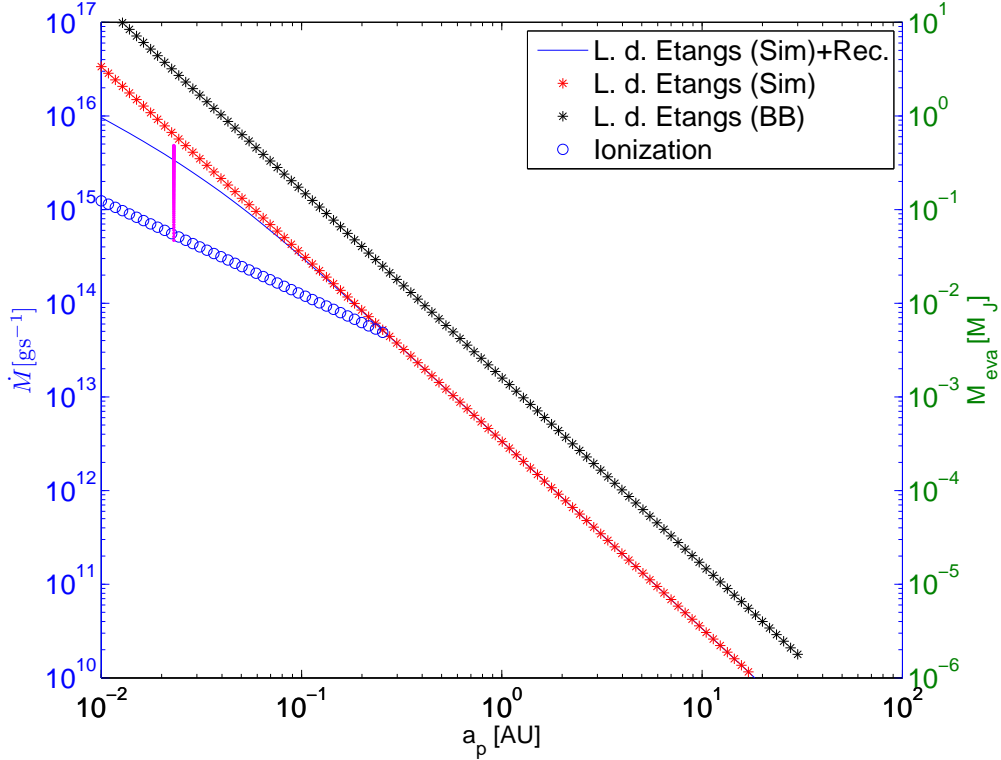}
\caption{Mass evaporation rate $\dot m_p$ (left axis) versus the
orbital separation $a_p$. The right axis gives the total mass that
would be evaporate during a period of $6\times 10^7{\rm yr}$. The
blue circles (lower line) represent the ionization model (for
details see Bear \& Soker 2010c). The black thick (upper) line
represents the evaporation rate based on Lecavelier des Etangs
(2007) for a black body energy distribution.The red thick line
represents the evaporation rate based on Lecavelier des Etangs
(2007) for a self consistent calculated spectrum of HD 149382
(Heber 2010). The blue thin line represents the same model of
Lecavelier des Etangs (2007) with recombination of the evaporated
gas included (eq. \ref{eq:dotmp4}) for a self consistent
calculated spectrum of HD 149382 (Heber 2010), instead of a black
body that is not accurate at wavelengths below 1200A. The
evaporation rates are calculated for an EHB central star and a
planet with the properties of the HD 149382 system: $T_{\rm
EHB}=35500K$, $M_{\rm EHB}=0.5M_\odot$, $R_{\rm EHB}=0.14R_\odot$,
$M_p=15M_{\rm J}$ (Geier et al. 2009) and $R_p=0.1R_\odot$. The
orbital separation of this system is $a_p=5-6.1R_\odot$, but here
it is an independent variable. The magenta line represents an
orbital separation of $a_p=5.5 R_\odot$.}\label{ML_180710}
\end{figure}

The recombination rate is proportional to the density square,
hence to mass loss square. We solve the mass loss rate of
Lecavelier des Etangs (2007) taking into account the
recombination. Substituting numerical values gives
\begin{equation}
\dot m_{p0} >  5\times 10^{15} \left( \frac{\beta}{0.5} \right)
\left(\frac {v_{\rm esc}} {250 {\rm km ~ s}^{-1}} \right)^4 \left(
\frac{R_p}{0.1 R_\odot} \right) \left( \frac{e_\gamma}{20 eV}
\right)^{-1} g s^{-1}. \label{eq:dotmp4}
\end{equation}

Where $4 \pi \beta$ is the solid angle to where the evaporation
flow occurs with $\beta=0.5$, $v_{\rm esc}$ is the escape velocity
of the gas, $R_p$ is the planet radius and $e_\gamma \sim 20 {\rm
eV}$ is the average energy of the ionizing photons. Figure
\ref{ML_180710} represents the different mass loss considered and
the mass loss that takes into account recombination. These are
calculated with the appropriate spectrum as was calculated for HD
149382 (Heber 2010). For comparison we show the evaporation rate
for a BB (Black Body) spectrum with the same effective temperature
and luminosity (black upper line, for details see figure caption).

The properties of the EHB central star and the planet are taken to
be those of the HD 149382 system (Geier et al. 2009; see figure
caption). The orbital separation of this system is
$a_p=5-6.1R_\odot$, but in the figure this is an independent
variable. On the right axis of figure \ref{ML_180710} we give the
total mass that would be evaporate during a period of $6\times
10^7{\rm yr}$, about the duration of the HB phase, with the same
mass loss rate given on the left axis. We will concentrate on the
orbital separation range of $a_p=0.01- 5AU$.

The calculation of the H$\alpha$ luminosity from the evaporated
gas is done in the following way, starting with the following
assumptions:
\begin{enumerate}
\item The evaporation is mainly into a solid angle $4 \pi \beta$.
\item Close to the planet, where most of the recombination occurs,
the material flows at the escape speed from the planet. \item For
typical values we find the medium to be optically thin to
H$\alpha$. \item We assume that the evaporated gas is almost
completely ionized. Any recombination that occurs is balanced by
the incoming photons from the EHB star. \item Most of the
recombination and the $H_\alpha$ source occur at a relatively high
density of $n \simeq 10^{10}-10^{12} {\rm cm^{-3}}$. At such
densities collision between atoms change the amount of energy that
is channelled to $H_\alpha$. In our simple treatment we take the
recombination coefficient neglecting the dependence on density. We
note that Bhatt (1985) calculates the H$\alpha$ emission from a
destructed comet. He estimates the density to be $\sim 10^{13}{\rm
cm^{-3}}$ and neglects the dependence on density. Korista et al.
(1997) found that the dependence in density on this high densities
is negligible.
\end{enumerate}

The H$\alpha$ energy released due to recombination is
\begin{equation}
L_{\rm H\alpha} = \int_{R_p}^{\infty}\alpha_H (h\nu_{H\alpha})n_e
n_p dV \label{Eq.Ha1}
\end{equation}
Solving the integral yields
\begin{equation}
L_{\rm H\alpha} \sim 3 \times 10^{29}   
\left(\frac{\dot M}{10^{16}{\rm g ~ s}^{-1}}\right)^2 \left(
\frac{\beta }{0.5} \right)^{-1}
\left(\frac{R_p}{0.1R_\odot}\right)^{-1} \left(\frac{v_{\rm
esc}}{250 {\rm km ~ s}^{-1}}\right)^{-2} {\rm erg ~ s}^{-1}.
\label{Eq.Ha2}
\end{equation}

We find equivalent width of $EW_{\alpha} \sim 0.05A$ for the
$H\alpha$ emission and  $EW_{\beta} \sim 0.006A$ for the $H\beta$
emission, both for the calculated spectrum of (Heber 2010).

Although the EWs are not high, their periodic variation might ease
the detection of the line. At an orbital separation of $5.5
R_\odot$ the orbital velocity of the substellar companion is $130
{\rm km ~ s}^{-1}$. Therefore, during the orbital period the
center of the emission by the evaporated gas might move back and
forth over a range of up to $\sim 5.5A$ and $\sim 4.0A$, for the
$H\alpha$ and $H\beta$ emission lines, respectively. These EWs are
an upper value since they are based on the black body
distribution. We conclude that it might be possible to identify a
planet via the H$\alpha$ emission of its ablated envelope.

\section{Summary and Conclusions}
\label{sec:Sec4}

We find that the pre CE evolution is crucial in understanding
binary systems where the primary star is an evolved red giant
branch (RGB) star, while the secondary star is a low-mass main
sequence (MS) star or a brown dwarf. Moving to planets, from
studying the distribution of planets we see that they are expected
in a double peak distribution. Outer planets can survive the
evolution of the progenitor of the EHB. In particular, inner
planets that were engulfed by the RGB progenitor might ``saved''
the outer planets by enhancing mass loss rate early on the RGB.
With the enhanced mass loss rate the RGB star will form an EHB
star. We also saw in section 3 that planets close to an EHB star
might be detected through H$\alpha$ and H$\beta$ emission.




\section*{ACKNOWLEDGMENTS}
This research was supported by the Asher Fund for Space Research
at the Technion, and the Israel Science foundation. E.B. was
supported in part by the Center for Absorption in Science,
Ministry of Immigrant Absorption, State of Israel.

\end{document}